\documentclass[aps,prb,twocolumn,superscriptaddress,longbibliography,10pt]{revtex4-1}
\usepackage{amssymb, amsmath, bm}
\usepackage{graphicx,onlyamsmath}

\usepackage{natbib}
\usepackage{xcolor}
\usepackage[normalem]{ulem}  
\usepackage{makecell} 

\begin{document}

\title{Quasiparticle picture of topological phase transitions induced by interactions}

\author{S. S.~Krishtopenko}
\affiliation{Laboratoire Charles Coulomb (L2C), UMR 5221 CNRS-Universit\'{e} de Montpellier, F-34095 Montpellier, France}
\affiliation{Julius-Maximilians-Universit\"{a}t W\"{u}rzburg, Physikalisches Institut and W\"{u}rzburg-Dresden Cluster of Excellence ct.qmat, Lehrstuhl f\"{u}r Technische Physik, Am Hubland, 97074 W\"{u}rzburg, Deutschland}

\author{A. V.~Ikonnikov}
\affiliation{Physics Department, M.V. Lomonosov Moscow State University, Moscow, 119991, Russia}

\author{F.~Hartmann}
\affiliation{Julius-Maximilians-Universit\"{a}t W\"{u}rzburg, Physikalisches Institut and W\"{u}rzburg-Dresden Cluster of Excellence ct.qmat, Lehrstuhl f\"{u}r Technische Physik, Am Hubland, 97074 W\"{u}rzburg, Deutschland}

\author{S.~H\"ofling}
\affiliation{Julius-Maximilians-Universit\"{a}t W\"{u}rzburg, Physikalisches Institut and W\"{u}rzburg-Dresden Cluster of Excellence ct.qmat, Lehrstuhl f\"{u}r Technische Physik, Am Hubland, 97074 W\"{u}rzburg, Deutschland}

\author{B.~Jouault}
\affiliation{Laboratoire Charles Coulomb (L2C), UMR 5221 CNRS-Universit\'{e} de Montpellier, F-34095 Montpellier, France}

\author{F.~Teppe}
\email[]{frederic.teppe@umontpellier.fr}
\affiliation{Laboratoire Charles Coulomb (L2C), UMR 5221 CNRS-Universit\'{e} de Montpellier, F-34095 Montpellier, France}


\begin{abstract}
We present a general recipe to describe topological phase transitions in condensed matter systems with interactions. We show that topological invariants in the presence of interactions can be efficiently calculated by means of a non-Hermitian quasiparticle Hamiltonian introduced on the basis of the Green's function. As an example analytically illustrating the application of the quasiparticle concept, we consider a topological phase transition induced by the short-range electrostatic disorder in a two dimensional system described by the Bernevig-Hughes-Zhang model. The latter allows us to explicitly demonstrate the change in the $\mathbb{Z}_2$ topological invariant and explain the quantized values of the longitudinal conductance in a certain range of the Fermi energy and the disorder strength found previously in numerical calculations.
\end{abstract}

\maketitle

\section{\label{Sec:Int} Introduction}
Topological insulators (TIs) are quantum states of matter with an insulating bulk and stable metallic surface states~\cite{QP1,QP2}. The origin of their physics is the bulk topology in momentum space, which can generally be characterized by topological invariants, the simplest examples among which are the TKNN~\cite{QP3} and the $\mathbb{Z}_2$ topological invariants~\cite{QP4,QP5,QP6,QP7,QP8,QP9,QP10}. The topological band invariants are defined in terms of the Bloch states, therefore, strictly speaking they can only apply to non-interacting systems, although one may extrapolate them to weakly interacting systems. A natural question is then whether adding interactions to TIs preserves their observable properties, such as the existence and robustness of the edge states. Moreover, the interactions itself may induce the topological phase transition into a TI state~\cite{QP11,QP12,QP13,QP14,QP15,QP16,QP17,QP18a,QP18b,QP18c,QP19}. It is therefore of prime importance to know how to correctly characterize topological systems in the presence of interactions.

One of the ways to take into account the interaction effect in the calculations of various topological invariants is the Green's function formalism~\cite{QP20,QP21,QP22,QP23,QP24,QP25,QP25b,QP25v1,QP25v2,QP25v3,QP26,QP27,QP28}. Within this formalism, a certain quantity characterizing the topology of the system, is expressed in terms of the single-particle retarded Green's function
\begin{equation}
\label{eq:1}
\hat{G}(\mathbf{k},\varepsilon)=
\left[\varepsilon-H_{0}(\mathbf{k})-\hat{\Sigma}(\mathbf{k},\varepsilon)\right]^{-1},
\end{equation}
where $H_{0}(\mathbf{k})$ is the non-interacting Hamiltonian (here, $\mathbf{k}$ is the quasimomentum, and $\varepsilon$ is considered as an external parameter, e.g. the Fermi energy), while the self-energy $\hat{\Sigma}(\mathbf{k},\varepsilon)$ incorporates the interaction effect. The latter includes all types of single-particle (electron-impurity, electron-phonon, etc.) and many-particle interactions that can be treated within the framework of single-particle Green's function approach. However, direct calculation of some topological invariants, such as the $\mathbb{Z}_2$ invariant, can be quite cumbersome, since it requires not only knowledge of the Green's function in the entire energy domain, but also the calculation of multifold integrals~\cite{QP21,QP22,QP23,QP24,QP25,QP25b,QP25v1,QP25v2,QP25v3}.

Z.~Wang and collaborators~\cite{QP26,QP27,QP28} proposed a recipe for the evaluation of a topological invariants in the presence of inversion symmetry based on the zero-energy Green's function. Their main idea is based on the introduction of ``topological Hamiltonian'' as
\begin{equation}
\label{eq:2}
H_{t}(\mathbf{k})\equiv-\hat{G}^{-1}(\mathbf{k},0)=H_{0}(\mathbf{k})+\hat{\Sigma}(\mathbf{k},0),
\end{equation}
which can be used for the calculations of topological invariants in the presence of interactions. Note that the Hamiltonian introduced in this way stays Hermitian~\cite{QP26,QP27,QP28,QP28a}, therefore topological characterization of the system with interactions is equivalent to the characterization of a ``non-interacting system'' with free Hamiltonian $H_{t}(\mathbf{k})$. Although the formalism based on the zero-energy Green's functions indeed often leads to correct topological characterization of different systems in the presence of single-particle and many-particle interactions~\cite{QP28,QP29,QP30,QP31,QP32,QP33,QP34}, there are still many cases of its breakdown~\cite{QP35}. The latter is apparently due to the fact that, unlike $\hat{G}(\mathbf{k},\varepsilon)$, $H_{t}(\mathbf{k})$ does not describe the system in a full manner.

Alternatively, we introduce the \emph{quasiparticle} Hamiltonian
\begin{equation}
\label{eq:3}
\mathcal{H}_{qp}(\mathbf{k},\varepsilon)=
H_{0}(\mathbf{k})+\hat{\Sigma}(\mathbf{k},\varepsilon),
\end{equation}
which, like the Green's function, contains all the information about the system. In the most general case, $\mathcal{H}_{qp}(\mathbf{k},\varepsilon)$ is \emph{non-Hermitian} and therefore has complex eigenvalues. Physically, the real part of its eigenvalues characterizes the ``renormalized'' band structure, while the imaginary part describes the quasiparticle decay resulting in the finite quasiparticle lifetime. The latter however is not crucial, since the methods developed in recent years~\cite{QP36o,QP36,QP36b,QP36c,QP37,QP38,QP39,QP40,QP41,QP42} allow topological characterization of non-Hermitian Hamiltonians as well. Since non-Hermitian systems have a richer phase diagram compared to Hermitian ones~\cite{QP36o,QP36,QP36b,QP36c,QP37,QP38}, the topological characterization of systems with interactions in the most general way should be performed by means of the quasiparticle Hamiltonian $\mathcal{H}_{qp}(\mathbf{k},\varepsilon)$ rather than $H_{t}(\mathbf{k})$.

The aim of this paper is to demonstrate an application of the quasiparticle description for interaction-induced phase transitions. For clarity, we focus on the calculation of the $\mathbb{Z}_2$ invariant in two-dimensional (2D) system described by the simplest two-band Bernevig-Hughes-Zhang (BHZ) model~\cite{QP43}. The latter characterizes almost all known time-reversal-invariant 2D TIs, including single HgTe/CdTe and
three-layer InAs/Ga(In)Sb quantum wells (QWs)~\cite{QP44,QP44b,QP44c,QP44d,QP45,QP46,QP47,QP47b,QP47c,QP48,QP49,QP50,QP50b,QP51,QP51b}. For the interaction, we consider a short-range disorder treated within the self-consistent Born approximation (SCBA)~\cite{QP52}.

Being added to initially trivial HgTe/CdTe QWs, the short-range disorder results in a non-trivial state (also called as ``topological Anderson insulator'') with quantized conductance values~\cite{QP16,QP17}. In our work, we not only analytically demonstrate the $\mathbb{Z}_2$ invariant change in the presence of the disorder, but also explain the quantized conductance values in a certain range of the Fermi energy and the disorder strength found previously in numerical calculations~\cite{QP16,QP17}. Our results provide insight into the application of the quasiparticle concept to describe various topological phase transitions and calculations of corresponding topological invariants in the presence of interactions.

\section{\label{Sec:A} BHZ model with interactions}

\subsection{\label{Sec:A2} Quasiparticle Hamiltonian and $\mathbb{Z}_2$ invariant}
Within the following sequence of the basis states $|E1{\uparrow}\rangle$, $|H1{\uparrow}\rangle$, $|E1{\downarrow}\rangle$, $|H1{\downarrow}\rangle$, the simplest version of BHZ Hamiltonian, ignoring the terms that break inversion and axial rotation symmetry, has the form
\begin{equation}
\label{eqA2:1}
\mathcal{H}_{0}(k_x, k_y)=\begin{pmatrix}
H_{\mathrm{BHZ}}(\mathbf{k}) & 0 \\ 0 & H_{\mathrm{BHZ}}^{*}(-\mathbf{k})\end{pmatrix},
\end{equation}
where the asterisk stands for complex conjugation, $\textbf{k}=(k_x,k_y)$ is the momentum in the plane, and
\begin{equation}
\label{eqA2:2}
H_{\mathrm{BHZ}}(\mathbf{k})=d_{0}(k)+\mathbf{d}(\mathbf{k})\cdot\vec{\sigma}.
\end{equation}
Here, $\vec{\sigma}=(\sigma_x,\sigma_y,\sigma_z)$ consists of the Pauli matrices; $d_{0}(k)=C-Dk^2$ and $\mathbf{d}=(d_x,d_y,d_z)$, where $d_x(\mathbf{k})=Ak_x$, $d_y(\mathbf{k})=Ak_y$, $d_z(k)=M-Bk^2$ and $k^2=k_x^2+k_y^2$. In Eq.~(\ref{eqA2:2}), the mass parameter $M$ describes inversion between the electron-like \emph{E}1 and hole-like \emph{H}1 subbands: $M>0$ and $M<0$ correspond to the trivial and quantum spin Hall insulator (QSHI) state, respectively~\cite{QP43}.The other structure parameters $A$, $B$, $C$, $D$ involved in $\mathcal{H}_{0}(\mathbf{k})$ depend on the QW growth direction, QW geometry and external conditions (such as temperature~\cite{QP46} or hydrostatic pressure~\cite{QP45b}).

For simplicity, without specifying the type of interaction preserving time reversal symmetry, we will assume that it results in the self-energy matrix $\hat{\Sigma}(\mathbf{k},\varepsilon)$, also of block-diagonal form:
\begin{equation}
\label{eqA2:3}
\hat{\Sigma}(\mathbf{k},\varepsilon)=\begin{pmatrix}
\Sigma^{\uparrow}(\mathbf{k},\varepsilon) & 0 \\ 0 & \Sigma^{\downarrow}(\mathbf{k},\varepsilon)\end{pmatrix}.
\end{equation}
Thus, the \emph{quasiparticle} Hamiltonian also has a block-diagonal form:
\begin{multline}
\label{eqA2:4}
\mathcal{H}_{qp}(\mathbf{k},\varepsilon)=\begin{pmatrix}
\mathcal{H}_{qp}^{\uparrow}(\mathbf{k},\varepsilon) & 0 \\ 0 & \mathcal{H}_{qp}^{\downarrow}(\mathbf{k},\varepsilon) \end{pmatrix}\\
=\mathcal{H}_{0}(\mathbf{k})+\begin{pmatrix}
\Sigma^{\uparrow}(\mathbf{k},\varepsilon) & 0 \\ 0 & \Sigma^{\downarrow}(\mathbf{k},\varepsilon)\end{pmatrix}.
\end{multline}

A block-diagonal form of $\mathcal{H}_{qp}(\mathbf{k},\varepsilon)$ reduces the calculation of the $\mathbb{Z}_2$ invariant to the calculation of
\begin{equation}
\label{eqA2:5}
\mathbb{Z}_2(\varepsilon)=\mathrm{mod}\left(\dfrac{C_{\uparrow}(\varepsilon)-C_{\downarrow}(\varepsilon)}{2},2\right),
\end{equation}
where $C_{\uparrow}(\varepsilon)$ and $C_{\downarrow}(\varepsilon)$ are the Chern numbers for the upper $\mathcal{H}_{qp}^{\uparrow}(\mathbf{k},\varepsilon)$ and lower $\mathcal{H}_{qp}^{\downarrow}(\mathbf{k},\varepsilon)$ blocks, respectively. We remind that $\varepsilon$ in Eq.~(\ref{eqA2:5}) should be considered as an external parameter, e.g. the Fermi energy. Since $C_{\uparrow}(\varepsilon)+C_{\downarrow}(\varepsilon)=0$ also holds for non-Hermitian systems~\cite{QP25b,QP53,QP54} due to the presence of time-reversal symmetry, we will further focus only on the calculation for the upper block $\mathcal{H}_{qp}^{\uparrow}(\mathbf{k},\varepsilon)$.

So far as $\mathcal{H}_{qp}^{\uparrow}(\mathbf{k},\varepsilon)$ is non-Hermitian, their left and right eigenstates are generally unrelated and satisfy the following eigenvalue equations:
\begin{eqnarray}
\label{eqA2:6}
\mathcal{H}_{qp}^{\uparrow}(\mathbf{k},\varepsilon)|\Psi_s^{R}(\mathbf{k},\varepsilon)\rangle=
E^{\uparrow}_s(\mathbf{k},\varepsilon)|\Psi_s^{R}(\mathbf{k},\varepsilon)\rangle,\notag\\
\left(\mathcal{H}_{qp}^{\uparrow}\right)^{\dag}(\mathbf{k},\varepsilon)|\Psi_s^{L}(\mathbf{k},\varepsilon)\rangle=
{E^{\uparrow}_s}^{*}(\mathbf{k},\varepsilon)|\Psi_s^{L}(\mathbf{k},\varepsilon)\rangle,
\end{eqnarray}
where $s$ is the band index that labels different eigenstates. If the system consists of only separable bands, such that $E^{\uparrow}_s(\mathbf{k},\varepsilon){\neq}E^{\uparrow}_l(\mathbf{k},\varepsilon)$ for all $s{\neq}l$ and all momentum $\mathbf{k}$, for any separable band with energy $E^{\uparrow}_s(\mathbf{k},\varepsilon)$, one can construct four different gauge invariant Berry curvatures~\cite{QP37}:
\begin{equation}
\label{eqA2:7}
B_{s,ij}^{\alpha\beta}(\mathbf{k},\varepsilon)=i\langle\partial_{i}\Psi_s^{\alpha}(\mathbf{k},\varepsilon)|\partial_{j}\Psi_s^{\beta}(\mathbf{k},\varepsilon)\rangle
\end{equation}
with the normalization condition $\langle\Psi_s^{\alpha}|\Psi_s^{\beta}\rangle=1$, where $i,j\in\left\{k_x,k_y\right\}$ and $\alpha,\beta\in\left\{L,R\right\}$. Importantly, although these  ``left-left'', ``left-right'', ``right-left'' and ``right-right'' Berry curvatures are locally different complex quantities, their integrals all yield the same real and quantized Chern number~\cite{QP37}:
\begin{equation}
\label{eqA2:8}
C^{(s)}_{\uparrow}(\varepsilon)=\dfrac{1}{2\pi}{\int}\epsilon_{ij}B_{s,ij}^{\alpha\beta}(\mathbf{k},\varepsilon)
{d^2\textbf{k}},
\end{equation}
where $\epsilon_{ij}=-\epsilon_{ji}$ denotes the Levi-Civita symbol in two dimensions and the summation over $i$ and $j$ is implied. After some transformations, Eq.~(\ref{eqA2:8}) can be also rewritten in vectorial notation:
\begin{equation}
\label{eqA2:9}
C^{(s)}_{\uparrow}(\varepsilon)=\dfrac{i}{2\pi}{\int}
\nabla_{\mathbf{k}}\times\langle\Psi_s^{L}(\mathbf{k},\varepsilon)|\nabla_{\mathbf{k}}\Psi_s^{R}(\mathbf{k},\varepsilon)\rangle
\cdot{\hat{z}_0}{d^2\textbf{k}}.
\end{equation}

To proceed further, we represent $\mathcal{H}_{qp}^{\uparrow}(\mathbf{k},\varepsilon)$ in the form similar to Eq.~(\ref{eqA2:2}):
\begin{equation}
\label{eqA2:10}
\mathcal{H}_{qp}^{\uparrow}(\mathbf{k},\varepsilon)=h_{0}(\mathbf{k},\varepsilon)+\mathbf{h}(\mathbf{k},\varepsilon)\cdot\vec{\sigma},
\end{equation}
where $h_{0}$ and $\mathbf{h}=(h_x,h_y,h_z)$ are \emph{complex} functions of $\mathbf{k}$ and $\varepsilon$. In this case, the Chern number for a \emph{complex} band with a given index $s$ is calculated as follows (see Appendix~\ref{sec:AppChern} for details):
\begin{equation}
\label{eqA2:11}
C^{(s)}_{\uparrow}(\varepsilon)=-\dfrac{s}{4\pi}{\int}
\dfrac{\mathbf{h}(\mathbf{k},\varepsilon)}{\lambda(\mathbf{k},\varepsilon)^3}\cdot\left[\nabla_{k_x}\mathbf{h}(\mathbf{k},\varepsilon)\times
\nabla_{k_y}\mathbf{h}(\mathbf{k},\varepsilon)\right]
{d^2\textbf{k}},
\end{equation}
where $s={\pm}1$ and $\lambda(\mathbf{k},\varepsilon)$ is defined as
\begin{equation}
\label{eqA2:12}
\lambda(\mathbf{k},\varepsilon)=\sqrt{h_x^2(\mathbf{k},\varepsilon)+h_y^2(\mathbf{k},\varepsilon)+h_z^2(\mathbf{k},\varepsilon)}.
\end{equation}


Finally, we note that the separability of complex bands, necessary for calculating the Chern numbers and the $\mathbb{Z}_2$ invariant~\cite{QP37}, is less stringent requirement than the presence of the band-gap in the quasiparticle spectrum. The latter is defined by the vanishing of the spectral function $A(k,\varepsilon)$ and density-of-states $D(\varepsilon)$:
\begin{eqnarray}
\label{eqA2:13}
A(\mathbf{k},\varepsilon)=-\dfrac{1}{\pi}\sum_{\sigma,s}\dfrac{\textrm{Im}E^{\sigma}_s(\mathbf{k},\varepsilon)}{\left(\varepsilon-\textrm{Re}E^{\sigma}_s(\mathbf{k},\varepsilon)\right)^2+\left(\textrm{Im}E^{\sigma}_s(\mathbf{k},\varepsilon)\right)^2},\notag\\
D(\varepsilon)={\int}A(\mathbf{k},\varepsilon){d^2\textbf{k}},~~~~~~~~~~~~~~~~~~~~~~
\end{eqnarray}
where an additional symbol $\sigma$ labels the eigenvalues from the upper and lower blocks of $\mathcal{H}_{qp}(\mathbf{k},\varepsilon)$. From Eq.~(\ref{eqA2:13}) it becomes clear that the condition of band separability ($E^{\sigma}_s(\mathbf{k},\varepsilon){\neq}E^{\sigma}_l(\mathbf{k},\varepsilon)$ for all $s{\neq}l$ and all $\mathbf{k}$) can also be satisfied out of the band-gap region or even in the absence of the band-gap.

\begin{table*}
\caption{\label{tab:1} Structure parameters of the prototype HgTe/Cd$_{0.7}$Hg$_{0.3}$Te QW used in the calculations.}
\begin{ruledtabular}
\begin{tabular}{cccccccc}
HgTe QW width (nm)  & Buffer & $a_0$~(nm) & $C$~(meV) & $M$~(meV) & $B$~(meV$\cdot$nm$^2$) & $D$~(meV$\cdot$nm$^2$) & $A$~(meV$\cdot$nm) \\
\hline
6.0  & (001) CdTe & 0.646 &  0 & 6.49 & -568.35 & -394.00 & 380.28
\end{tabular}
\end{ruledtabular}
\end{table*}

\subsection{\label{Sec:A1} Short-range electrostatic disorder and SCBA }
The quasiparticle concept, presented in the previous section, allows one to calculate the $\mathbb{Z}_2$ invariant for an arbitrary dependence of the self-energy matrix $\hat{\Sigma}(\mathbf{k},\varepsilon)$ on $\mathbf{k}$ and $\varepsilon$, that is determined by the specific type of interaction in the 2D system. The latter includes both single-particle (electron-impurity, electron-phonon, etc.) and many-particle interactions, which can be treated within the framework of the single-particle Green's functions.

In order to illustrate the quasiparticle approach for calculating the $\mathbb{Z}_2$ invariant, we further consider the interaction with a short-range electrostatic disorder~\cite{QP16,QP17}, which will be treated for simplicity within the SCBA. Our choice of this illustrative example, as will be seen later, is motivated by two factors: (i) the calculations of the $\mathbb{Z}_2$ invariant are performed analytically; (ii) the topological phase transition induced by short-range electrostatic disorder cannot be described within the framework of ``topological Hamiltonian'' $H_{t}(\mathbf{k})$ proposed by Z.~Wang and collaborators~\cite{QP26,QP27,QP28}.

To introduce disorder into the system, we add a diagonal random impurity potential to the Hamiltonian $\mathcal{H}_{0}(\mathbf{k})$ in Eq.~(\ref{eqA2:1}):
\begin{equation}
\label{eqA1:1}
V_{imp}(\textbf{r})=\sum_{j}v(\textbf{r}-\textbf{R}_j),~~~~ v(\textbf{r})=\int\dfrac{d^2\textbf{q}}{(2\pi)^2}\tilde{v}(\textbf{q})e^{i\textbf{q}\cdot\textbf{r}},
\end{equation}
where $R_j$ denotes position of impurities and $v(\textbf{r})$ is the potential of an individual impurity. The latter is assumed to be isotropic, i.e. $\tilde{v}(\textbf{q}) = \tilde{v}({q})$ with $|\textbf{q}| = q$.

Then, the disorder-averaged Green's function $\hat{G}(\mathbf{k},\varepsilon)$ is written as
\begin{equation}
\label{eqA1:2}
\hat{G}(\mathbf{k},\varepsilon)=\langle\dfrac{1}{\varepsilon-\mathcal{H}}\rangle=
\left[\varepsilon-\mathcal{H}_{0}(\mathbf{k})-\hat{\Sigma}(\mathbf{k},\varepsilon)\right]^{-1},
\end{equation}
with
\begin{equation}
\label{eqA1:3}
\mathcal{H}=\mathcal{H}_{0}(\mathbf{k})+V_{imp}(\textbf{r}),
\end{equation}
where $\langle...\rangle$ denotes average over all disorder configurations. Within the SCBA, the self-energy matrix has a form
\begin{equation}
\label{eqA1:4}
\hat{\Sigma}(\textbf{k},\varepsilon)=n_{i}\int\dfrac{d^2\textbf{k}^\prime}{(2\pi)^2}
\tilde{v}(\textbf{k}-\textbf{k}^\prime)\hat{G}(\mathbf{k}^\prime,\varepsilon)\tilde{v}(\textbf{k}^\prime-\textbf{k}),
\end{equation}
where $n_{i}$ is the concentration of impurities.

The axial rotation symmetry of $\mathcal{H}_{0}(k_x,k_y)$ allows one to reduce Eq.~(\ref{eqA1:4}) to the form (see Appendix~\ref{sec:AppRot} for details):
\begin{equation}
\label{eqA1:5}
\hat{\Sigma}(k,\varepsilon)=\begin{pmatrix}
\hat{\Sigma}^{\uparrow}(k,\varepsilon) & 0\\
0 & \hat{\Sigma}^{\downarrow}(k,\varepsilon) \end{pmatrix}
\end{equation}
with
\begin{eqnarray*}
\hat{\Sigma}^{\uparrow}(k,\varepsilon)=n_{i}\int\limits_0^{K_c}\dfrac{k^\prime dk^\prime}{2\pi}\begin{pmatrix}
V_0(k,k^\prime)^2G_{11}^\prime & V_{+1}(k,k^\prime)^2G_{12}^\prime\\
V_{-1}(k,k^\prime)^2G_{21}^\prime & V_0(k,k^\prime)^2G_{22}^\prime  \end{pmatrix},\nonumber\\
\hat{\Sigma}^{\downarrow}(k,\varepsilon)=n_{i}\int\limits_0^{K_c}\dfrac{k^\prime dk^\prime}{2\pi}\begin{pmatrix}
V_0(k,k^\prime)^2G_{33}^\prime & V_{-1}(k,k^\prime)^2G_{34}^\prime\\
V_{+1}(k,k^\prime)^2G_{43}^\prime & V_0(k,k^\prime)^2G_{44}^\prime  \end{pmatrix},
\end{eqnarray*}
where $k=|\textbf{k}|$, $k^\prime=|\textbf{k}^\prime|$,
\begin{equation}
\label{eqA1:6}
V_n(k,k^\prime)^2=\int\limits_0^{2\pi}\dfrac{d\theta}{2\pi}|\tilde{v}(\textbf{k}-\textbf{k}^\prime)|^2
\cos n\theta,
\end{equation}
and $G_{ij}^\prime\equiv G_{ij}(k^\prime,\varepsilon)$ are the component of the Green's function:
\begin{equation}
\label{eqA1:7}
\hat{G}(k,\varepsilon)=\left[\varepsilon-\tilde{\mathcal{H}}_{0}(k)-\hat{\Sigma}(k,\varepsilon)\right]^{-1},
\end{equation}
with $\tilde{\mathcal{H}}_{0}(k)\equiv\mathcal{H}_{0}(k,0)$ in Eq.~(\ref{eqA2:1}). In the equations above, we introduce the angle $\theta$ between $\mathbf{k}$ and $\textbf{k}^\prime$, as well as the cut-off wave-vector $K_c=\pi/a_{\mathrm{0}}$ (where $a_{\mathrm{0}}$ is the lattice constant in the plane of 2D system, see Tab.~\ref{tab:1}), which corresponds to the size of the first Brillouin zone.

\begin{figure*}
\includegraphics [width=0.96\textwidth, keepaspectratio] {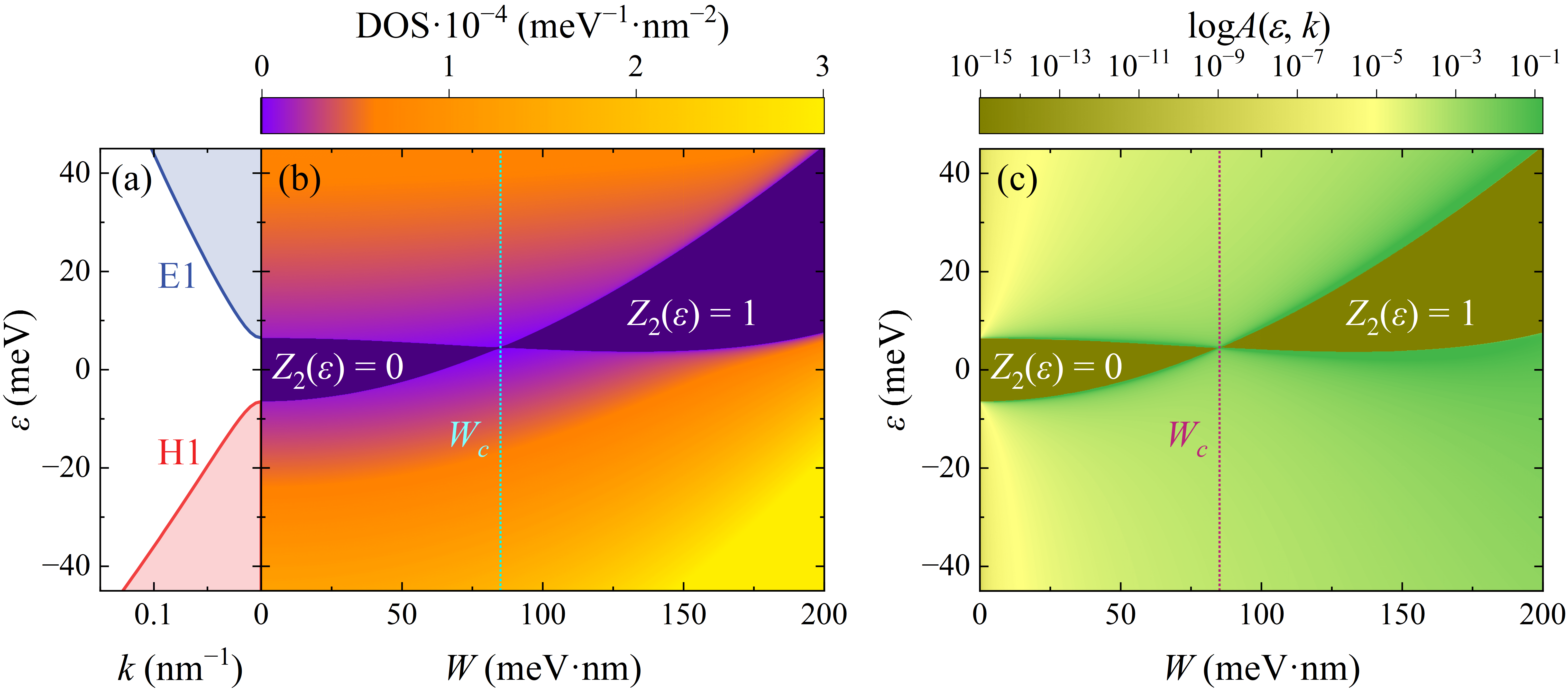} 
\caption{\label{Fig:1} (a) Low energy band structure of the prototype 6~nm wide HgTe/Cd$_{0.7}$Hg$_{0.3}$Te QW calculated on the basis of the BHZ model. (b) Color map of the density-of-states (DOS) and (c) the spectral function at the $\Gamma$ point as a function of the short-range disorder strength $W=\sqrt{{n}_{i}}u_0$ (see Eq.~(\ref{eqA1:8})). The band-gap region on $(\varepsilon,W)$ plane is represented by violet (in panel (b)) and olive (in panel (c)) area.}
\end{figure*}

As clear, Eqs.~(\ref{eqA1:5})--(\ref{eqA1:7}) form a system of integral equations that determines both the Green's function $\hat{G}(k,\varepsilon)$ and the self-energy $\hat{\Sigma}(\mathbf{k},\varepsilon)$ matrices. The self-consistent solution of such integral systems in the general case is a laborious task. However, for the case of the disorder formed by the short-range impurities, for which $\tilde{v}(q)=u_0$ (cf. Ref.~\cite{QP18c}), the solution of the problem can be significantly simplified.

Indeed, in the case of the short-range disorder, $V_n(k,k^\prime)^2=u_0^2\delta_{n,0}$ in Eq.~(\ref{eqA1:6}). The latter results in a diagonal form of the self-energy matrix $\hat{\Sigma}(\varepsilon)$ in Eq.~(\ref{eqA1:5}) being independent of the momentum. By means of direct calculations, one can verify that $G_{11}\left(k,\varepsilon\right)=G_{33}\left(k,\varepsilon\right)$ and $G_{22}\left(k,\varepsilon\right)=G_{44}\left(k,\varepsilon\right)$, which leads to
\begin{equation}
\label{eqA1:8}
\hat{\Sigma}^{\uparrow}(\varepsilon)=\hat{\Sigma}^{\downarrow}(\varepsilon)=
\dfrac{W^2}{4\pi}\int\limits_0^{K_c^2}dx
\begin{pmatrix}
G_{11}\left(\sqrt{x},\varepsilon\right) & 0 \\[4pt]
0 & G_{22}\left(\sqrt{x},\varepsilon\right)
\end{pmatrix},
\end{equation}
where $W^2=n_{i}u_0^2$ is the disorder strength. Note that integration in Eq.~(\ref{eqA1:8}) can be performed analytically, which transforms Eqs.~(\ref{eqA1:5})--(\ref{eqA1:8}) into the set of algebraic equations numerically solved by simple iterations.

Once the self-energy is known, the spectral function $A(k,\varepsilon)$ and density-of-states $D(\varepsilon)$ for the case of the short-range disorder can be found as
\begin{eqnarray}
\label{eqA1:9}
A(k,\varepsilon)=-\dfrac{2}{\pi}\textrm{Im}\left\{G_{11}\left(k,\varepsilon\right)+G_{22}\left(k,\varepsilon\right)\right\},\notag\\
D(\varepsilon)=\int\limits_0^{K_c}\dfrac{k dk}{2\pi}A(k,\varepsilon).~~~~~~~~~~~~
\end{eqnarray}

\section{\label{Sec:RnD} Results and discussion}
Figure~\ref{Fig:1} represents the evolution of density-of-states $D(\varepsilon)$ and spectral function $A(k=0,\varepsilon)$ at the $\Gamma$ point as a function of the short-range disorder strength $W=\sqrt{{n}_{i}}u_0$ calculated with the structure parameters of the prototype $6$~nm~wide HgTe/Cd$_{0.7}$Hg$_{0.3}$Te QW (see Tab.~\ref{tab:1}). All details about the calculation of structural parameters involved in $\mathcal{H}_{0}(\mathbf{k})$ can be found elsewhere~\cite{QP56}. As seen, being added to an initially clean trivial QW ($M>0$), the short-range disorder causes the band-gap to decrease with increasing $W$ until it vanishes at a critical value $W_c$ and then reopens again at $W>W_c$. Such band-gap behavior represents the disorder-induced topological phase transition previously discovered in the tight-binding calculations~\cite{QP16,QP17}. Importantly, the band-gap reopening region at $W>W_c$ was characterized by quantized conductance values inherent in the QSHI state. It is on the basis of this fact that the conclusion is made that $W=W_c$ corresponds to a topological phase transition.

\begin{figure*}
\includegraphics [width=\textwidth, keepaspectratio] {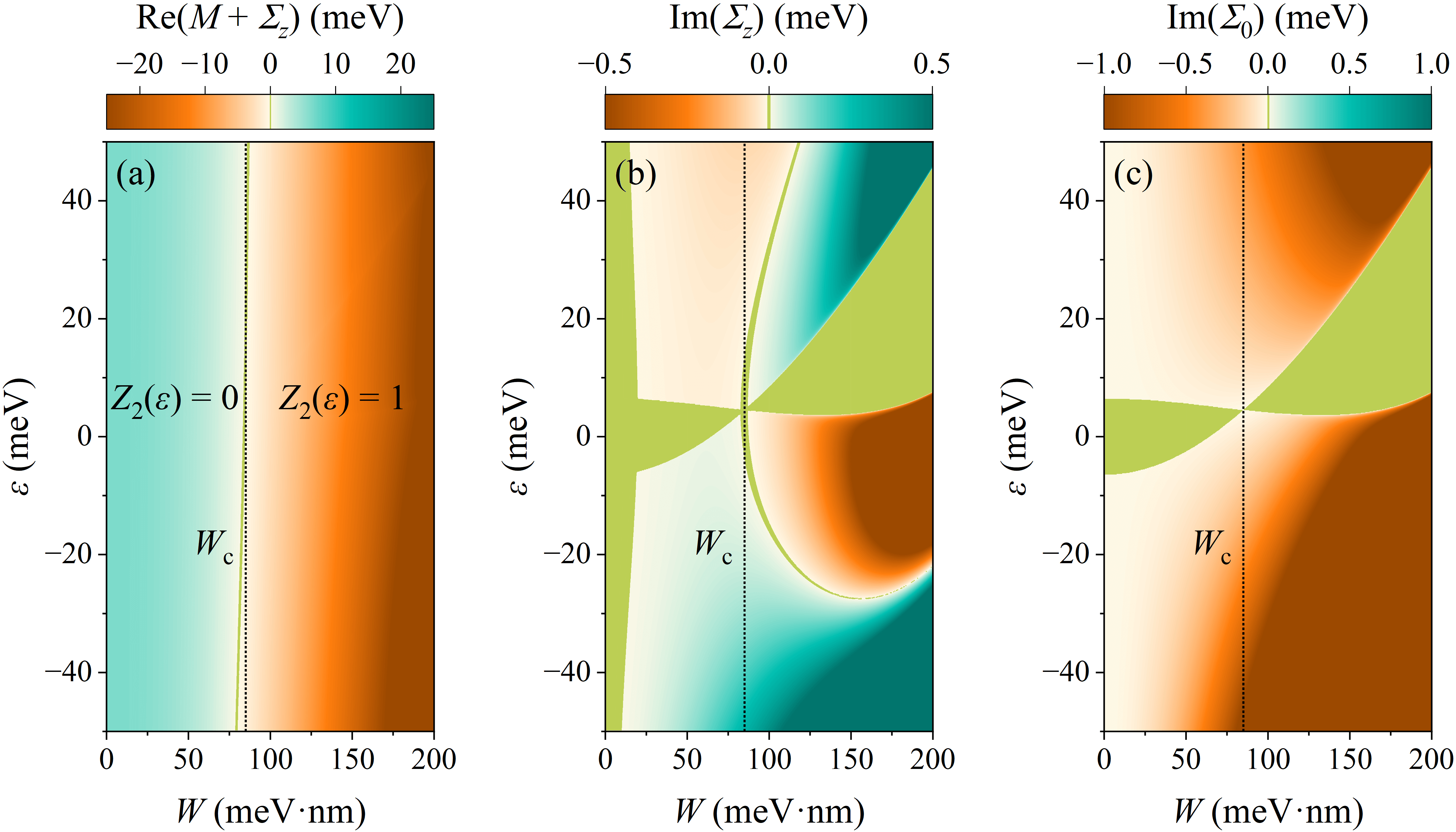} 
\caption{\label{Fig:2} Color map of the renormalized mass parameter $M+\textrm{Re}\Sigma_z(\varepsilon)$ (in panel (a)) and imaginary parts of $\Sigma_z(\varepsilon)$ (in panel (b)) and $\Sigma_0(\varepsilon)$ (in panel (x)) in the prototype 6~nm wide HgTe/Cd$_{0.7}$Hg$_{0.3}$Te QW as a function of the short-range disorder strength $W$. The zero-energy values are marked separately in green. The dotted vertical curves represent a critical value $W=W_c$.}
\end{figure*}

Let us now show that the band-gap behavior illustrated by the evolution of $D(\varepsilon)$ and $A(k=0,\varepsilon)$ is indeed due to a change in the $\mathbb{Z}_2$ invariant. For this purpose, it is convenient to represent the self-energy blocks in Eq.~(\ref{eqA1:8}) in the form
\begin{equation}
\label{eqRnD:1}
\hat{\Sigma}^{\uparrow}(\varepsilon)=\hat{\Sigma}^{\downarrow}(\varepsilon)
=\Sigma_0(\varepsilon)+\Sigma_z(\varepsilon)\sigma_z.
\end{equation}
Then, using the polar coordinate system, the Chern number for the valence band of $\mathcal{H}_{qp}^{\uparrow}(\mathbf{k},\varepsilon)$ can be calculated analytically
\begin{multline}
\label{eqRnD:2}
C^{(-)}_{\uparrow}(\varepsilon)=\dfrac{A^2}{4}\int\limits_{0}^{\infty}
\dfrac{Bx+M+\Sigma_z(\varepsilon)}{\left[A^2x+\left(M+\Sigma_z(\varepsilon)-Bx\right)^2\right]^{3/2}}dx\\
=\dfrac{B}{2\sqrt{B^2}}+\dfrac{M+\Sigma_z(\varepsilon)}{2\sqrt{\left(M+\Sigma_z(\varepsilon)\right)^2}}~\\
=\dfrac{\mathrm{sgn}\left[B\right]+\mathrm{sgn}\left[\textrm{Re}\left\{M+\Sigma_z(\varepsilon)\right\}\right]}{2}.
\end{multline}
For the latter, one should take into consideration Eq.~(\ref{eq:B4}) in Appendix~\ref{sec:AppChern}. Since parameter $B$ is always negative in HgTe QWs~\cite{QP45,QP46} (see also Tab.~\ref{tab:1}), then Eq.~(\ref{eqA2:5}) for the $\mathbb{Z}_2$ invariant becomes
\begin{equation}
\label{eqRnD:3}
\mathbb{Z}_2(\varepsilon)=\dfrac{1-\mathrm{sgn}\left[M+\textrm{Re}\Sigma_z(\varepsilon)\right]}{2}.
\end{equation}
Note that Eq.~(\ref{eqRnD:3}) is also valid at the energies $\varepsilon$ outside the band-gap region, where the imaginary part of both $\Sigma_0(\varepsilon)$ and $\Sigma_z(\varepsilon)$ is non-zero. The fact that the Chern numbers and the $\mathbb{Z}_2$ invariant above are independent of the imaginary part of the self-energy matrix is due primarily to the specificity of the interaction with isotropic short-range electrostatic impurities. In the most general case of interaction, the invariants can also depend on the imaginary part of the self-energy matrix (see Appendix~\ref{sec:AppSigma} for an example).

A parameter $M+\textrm{Re}\Sigma_z(\varepsilon)$ in Eq.~(\ref{eqRnD:3}) can be naturally treated as renormalized mass parameter -- a topological phase transition occurs when it changes its sign. Indeed, straightforward calculations presented in Fig.~\ref{Fig:2}(a) show that the band-gap closing at $W=W_c$ revealed by $D(\varepsilon)$ and $A(k=0,\varepsilon)$ clearly corresponds to zero values of $M+\textrm{Re}\Sigma_z(\varepsilon)$. Therefore, the band-gap region in Fig.~\ref{Fig:1} at $W<W_c$ is characterized by the positive values of $M+\textrm{Re}\Sigma_z(\varepsilon)$ and, hence, by $\mathbb{Z}_2=0$ in accordance with Eq.~(\ref{eqRnD:3}). On the contrary, the band-gap region at $W>W_c$ corresponds to $M+\textrm{Re}\Sigma_z(\varepsilon)<0$, resulting in $\mathbb{Z}_2=1$. In this way, the above calculation of $\mathbb{Z}_2$ topological invariant
fully confirms the intuitive conclusions about the role of $M+\textrm{Re}\Sigma_z(\varepsilon)$ in the topological phase transition made by Groth \emph{et al.}~\cite{QP17}.

As noted above, the disorder-induced topological phase transition was first discovered by means of numerical calculations of the longitudinal conductance in the presence of an external bias, which allowed the quantized values inherent in the QSHI state to be found out~\cite{QP16,QP17}. The latter is not obvious, since, for example, the Hall conductance of a non-Hermitian Chern insulator can deviate from its quantized value even if the Chern number is quantized~\cite{QP40,QP41}. Since our quasiparticle Hamiltonian (\ref{eqA2:4}) essentially consists of two copies of the Chern insulators described by $\mathcal{H}_{qp}^{\uparrow}(\mathbf{k},\varepsilon)$ and $\mathcal{H}_{qp}^{\downarrow}(\mathbf{k},\varepsilon)$, we next discuss the reasons why the disorder-induced phase with $\mathbb{Z}_2(\varepsilon)=1$ exhibits the quantized longitudinal conductance first observed by Li~\emph{et~al.}~\cite{QP16}.

Let us first focus on the Hall conductance due to the upper block $\mathcal{H}_{qp}^{\uparrow}(\mathbf{k},\varepsilon)$ of $\mathcal{H}_{qp}(\mathbf{k},\varepsilon)$ in Eq.~(\ref{eqA2:4}). To neglect the contribution from non-Hermitian bulk states, we further consider the range of $\varepsilon$ corresponding to the region defined as~\cite{QP40}
\begin{equation}
\label{eqRnD:4}
\textrm{Re}E_{-}^{\uparrow}(0,\varepsilon)<\varepsilon<\textrm{Re}E_{+}^{\uparrow}(0,\varepsilon),
\end{equation}
where
\begin{equation}
\label{eqRnD:5}
E_{\pm}^{\uparrow}(\mathbf{k},\varepsilon)=h_{0}(\mathbf{k},\varepsilon)\pm{\lambda(\mathbf{k},\varepsilon)}
\end{equation}
with $\lambda(\mathbf{k},\varepsilon)$ defined by Eq.~(\ref{eqA2:12}). Note that Eq.~(\ref{eqRnD:4}) obviously guarantees the fulfillment of the condition for the existence of separable bands in two-band non-Hermitian systems, which is necessary for calculating the Chern number (see Section~\ref{Sec:A2}). Taking into account Eq.~(\ref{eqRnD:4}), the Hall conductance caused by the upper spin block $\mathcal{H}_{qp}^{\uparrow}(\mathbf{k},\varepsilon)$ can be written as~\cite{QP40}:
\begin{equation}
\label{eqRnD:6}
\sigma_{xy}^{\uparrow}(\varepsilon)=\dfrac{1}{4\pi}\dfrac{e^2}{h}{\int}
\left[\dfrac{\Omega_{xy}(\mathbf{k},\varepsilon)+\Omega_{xy}^{*}(\mathbf{k},\varepsilon)}{2}\cdot
\nu(\mathbf{k},\varepsilon)\right]
{d^2\textbf{k}},
\end{equation}
where
\begin{multline}
\label{eqRnD:7}
\Omega_{xy}(\mathbf{k},\varepsilon)
=\dfrac{\mathbf{h}(\mathbf{k},\varepsilon)}{\lambda(\mathbf{k},\varepsilon)^3}\cdot\bigl[\nabla_{k_x}
\textrm{Re}\left\{\mathbf{h}(\mathbf{k},\varepsilon)\right\}\\
\times\nabla_{k_y}\textrm{Re}\left\{\mathbf{h}(\mathbf{k},\varepsilon)\right\}\bigr],
\end{multline}
and
\begin{multline}
\label{eqRnD:8}
\nu(\mathbf{k},\varepsilon)=\dfrac{1}{\pi}\biggl\{ \arctan\dfrac{\textrm{Re}\left\{h_{0}(\mathbf{k},\varepsilon)+\lambda(\mathbf{k},\varepsilon)-\varepsilon\right\}}
{\left|\textrm{Im}\left\{h_{0}(\mathbf{k},\varepsilon)+\lambda(\mathbf{k},\varepsilon)\right\}\right|}~~~~~~~~~~~\\
-\arctan\dfrac{\textrm{Re}\left\{h_{0}(\mathbf{k},\varepsilon)-\lambda(\mathbf{k},\varepsilon)-\varepsilon\right\}}
{\left|\textrm{Im}\left\{h_{0}(\mathbf{k},\varepsilon)-\lambda(\mathbf{k},\varepsilon)\right\}\right|}
\biggr\}.
\end{multline}
As seen, $\sigma_{xy}^{\uparrow}(\varepsilon)$ is not quantized in the most general case of non-Hermitian Hamiltonian, even if Chern number $C^{(-)}_{\uparrow}(\varepsilon)$ in Eq.~(\ref{eqA2:11}) is integer.

Figure~\ref{Fig:2}(b,c) provides the key to explain the quantized conductance values observed in the nontrivial phase induced by the short-range disorder~\cite{QP16,QP17}. As clearly seen, in the band-gap region identified by $D(\varepsilon)$ and $A(k=0,\varepsilon)$ in Fig.~\ref{Fig:1}, the imaginary part of $\Sigma_0(\varepsilon)$ and $\Sigma_z(\varepsilon)$ defined by Eq.~(\ref{eqRnD:1}) vanishes. This means that the quasiparticle Hamiltonian $\mathcal{H}_{qp}(\mathbf{k},\varepsilon)$ \emph{is Hermitian in the band-gap region}, and $\mathbf{h}(\mathbf{k},\varepsilon)$, $h_{0}(\mathbf{k},\varepsilon)$ and $\lambda(\mathbf{k},\varepsilon)$ become purely real functions. The latter leads to $\nu(\mathbf{k},\varepsilon)=1$ and
\begin{multline}
\label{eqRnD:9}
\sigma_{xy}^{\uparrow}(\varepsilon)=\dfrac{1}{4\pi}\dfrac{e^2}{h}{\int}
\dfrac{\mathbf{h}(\mathbf{k},\varepsilon)}{\lambda(\mathbf{k},\varepsilon)^3}\cdot\bigl[\nabla_{k_x}
\left\{\mathbf{h}(\mathbf{k},\varepsilon)\right\}~~~~~~~~~\\
\times\nabla_{k_y}\left\{\mathbf{h}(\mathbf{k},\varepsilon)\right\}\bigr]
{d^2\textbf{k}}=\dfrac{e^2}{h}C^{(-)}_{\uparrow}(\varepsilon),
\end{multline}
where $C^{(-)}_{\uparrow}(\varepsilon)$ has integer values in accordance with Eq.~(\ref{eqRnD:2}). By means of straightforward calculations for the lower spin block $\mathcal{H}_{qp}^{\downarrow}(\mathbf{k},\varepsilon)$, one can verify that inside the band-gap region
\begin{equation}
\label{eqRnD:10}
\sigma_{xy}^{\downarrow}(\varepsilon)=\dfrac{e^2}{h}C^{(-)}_{\downarrow}(\varepsilon)=-\dfrac{e^2}{h}C^{(-)}_{\uparrow}(\varepsilon),
\end{equation}
where $C^{(-)}_{\uparrow}(\varepsilon)+C^{(-)}_{\downarrow}(\varepsilon)=0$ due to the time-reversal symmetry.

Expressions~(\ref{eqRnD:9}) and (\ref{eqRnD:10}) show that, in the absence of external bias, the total Hall conductance $\sigma_{xy}^{\downarrow}(\varepsilon)+\sigma_{xy}^{\uparrow}(\varepsilon)$ in the band-gap region vanishes, while the spin Hall conductance in the band-gap region $[\sigma_{xy}^{\downarrow}(\varepsilon)-\sigma_{xy}^{\uparrow}(\varepsilon)]/2={e^2}/{h}$ if $\mathbb{Z}_2(\varepsilon)=1$, just like for the QSHI state in the clean limit~\cite{QP43}. The latter implies the existence of a pair of helical edge states inside the band-gap region shown in Fig.~\ref{Fig:1} at $W>W_c$.

\begin{figure*}
\includegraphics [width=0.9\textwidth, keepaspectratio] {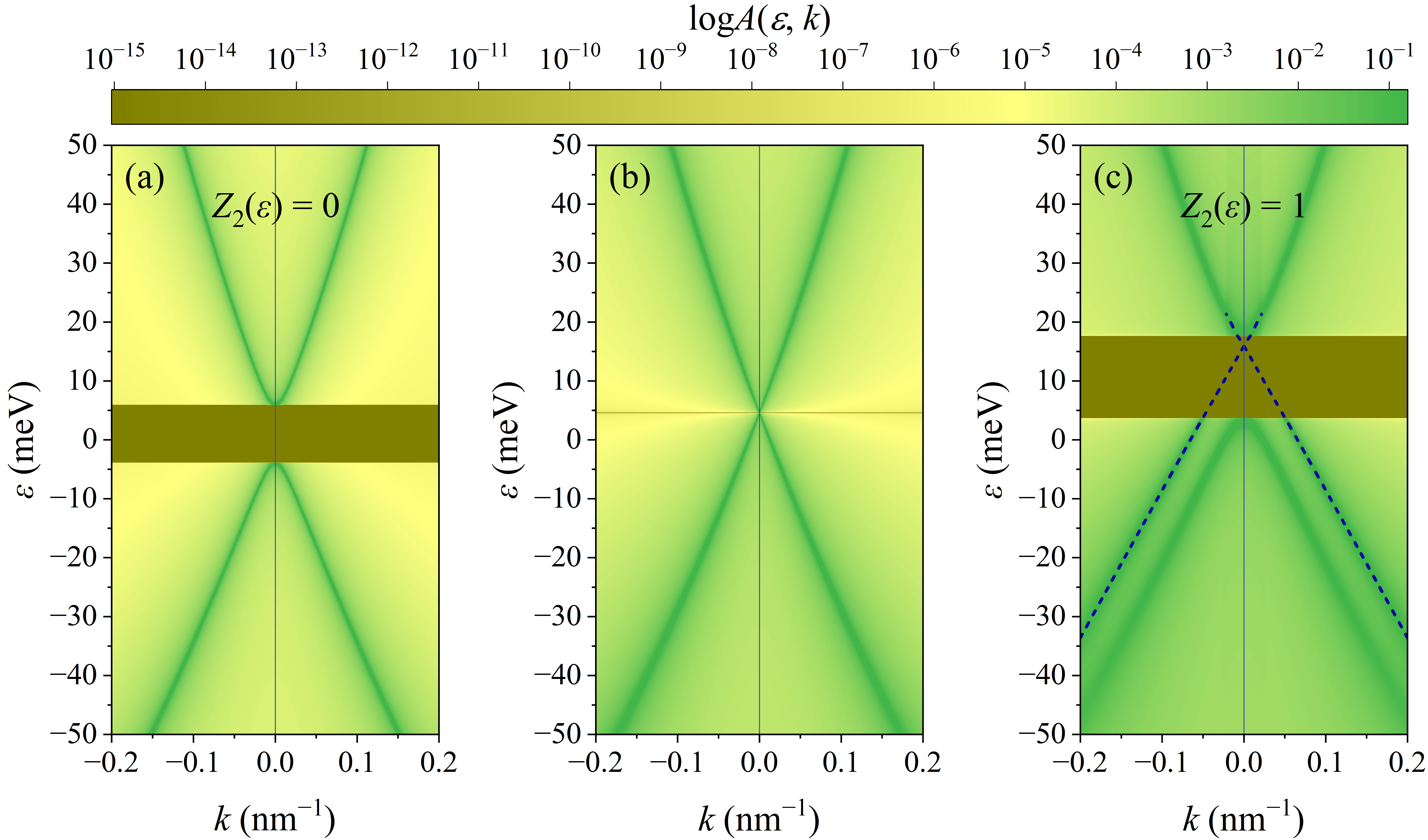} 
\caption{\label{Fig:3} Total spectral function $A(k,\varepsilon)+A^{\mathrm{edge}}(k,\varepsilon)$ determined by Eqs~(\ref{eqA1:9}) and (\ref{eqRnD:15}) in the prototype 6~nm wide HgTe/Cd$_{0.7}$Hg$_{0.3}$Te QW at several values of the short-range disorder strength: (a) $W = 40$~meV$\cdot$nm, (b) $W = W_c$, (c) $W = 130$~meV$\cdot$nm. The dashed blue lines in panel (c) correspond to the quasiparticle edge dispersion, determined by the maximum at $A^{\mathrm{edge}}(k,\varepsilon)$. Note that $A^{\mathrm{edge}}(k,\varepsilon)$ has the form of a delta-function in the band-gap region, since the damping factor $\Gamma(\varepsilon)$ in Eq.~(\ref{eqRnD:16}) vanishes therein.}
\end{figure*}

For concreteness, let us show the existence of quasiparticle helical edge states in $\mathcal{H}_{qp}(\mathbf{k},\varepsilon)$ under open boundary condition. To consider the edge state on a single edge, we deal with a system on a half-plane of $y\geq0$. Then, the eigenvalues and eigenstates for the states localized in the vicinity of $y=0$ are found analytically similar to the Hermitian case~\cite{QP57}:
\begin{equation}
\label{eqRnD:11}
E_{\sigma}^{\mathrm{edge}}(k_x,\varepsilon)=\tilde{C}(\varepsilon)-\dfrac{D}{B}\tilde{M}(\varepsilon)
+\sigma\sqrt{1-\dfrac{D^2}{B^2}}Ak_x,
\end{equation}
and
\begin{equation}
\label{eqRnD:12}
|\Psi_{\sigma}^{\mathrm{edge}}(k_x,\varepsilon)\rangle=\dfrac{e^{ik_{x}x}}{\sqrt{L_x}}g_{\sigma}(k_x,y)\chi_{\sigma},
\end{equation}
where $\tilde{C}(\varepsilon)=C+\Sigma_0(\varepsilon)$ and $\tilde{M}(\varepsilon)=M+\Sigma_z(\varepsilon)$; $\sigma=+1$ and $\sigma=-1$ corresponds to $\mathcal{H}_{qp}^{\uparrow}(\mathbf{k},\varepsilon)$ and $\mathcal{H}_{qp}^{\downarrow}(\mathbf{k},\varepsilon)$, respectively; while
\begin{eqnarray}
\label{eqRnD:13}
g_{\sigma}(k_x,y)=\sqrt{\left|\dfrac{2\lambda_{1}^{\sigma}\lambda_{2}^{\sigma}\left(\lambda_{1}^{\sigma}+\lambda_{2}^{\sigma}\right)}{\left(\lambda_{1}^{\sigma}-\lambda_{2}^{\sigma}\right)^2}\right|}
\left(e^{-\lambda_{1}^{\sigma}y}-e^{-\lambda_{2}^{\sigma}y}\right),\notag\\
\chi_{+1}=\begin{pmatrix}
1 & \eta & 0 & 0\end{pmatrix}/\sqrt{1+\eta^2},~~~~~~~~~~~~\notag\\
\chi_{-1}=\begin{pmatrix}
0 & 0 & 1 & \eta \end{pmatrix}/\sqrt{1+\eta^2}~~~~~~~~~~~~~
\end{eqnarray}
with $\eta^2=(B+D)/(B-D)$ and
\begin{eqnarray}
\label{eqRnD:14}
\lambda_{1,2}^{\sigma}=\sqrt{k_x^2+F\pm\sqrt{F^2-\dfrac{\tilde{M}(\varepsilon)^2-\left(E_{\sigma}^{\mathrm{edge}}-\tilde{C}(\varepsilon)\right)^2}{B^2-D^2}}},\notag\\
F=\dfrac{A^2-2\left[\tilde{M}(\varepsilon)B+\left(E_{\sigma}^{\mathrm{edge}}-\tilde{C}(\varepsilon)\right)D\right]}{2\left(B^2-D^2\right)}.~~~~~~~~
\end{eqnarray}
It follows from above that the existence condition for the helical edge states is the simultaneous fulfillment of the conditions $\textrm{Re}\left[\lambda_{1}^{\sigma}\right]>0$ and $\textrm{Re}\left[\lambda_{2}^{\sigma}\right]>0$.

Once $E_{\sigma}^{\mathrm{edge}}(k_x,\varepsilon)$ is known, in accordance with Eq.~(\ref{eqA2:13}), the spectral function for the quasiparticle edge states can be written as
\begin{equation}
\label{eqRnD:15}
A^{\mathrm{edge}}(k_x,\varepsilon)=-\dfrac{1}{\pi}\sum_{\sigma}\dfrac{\Gamma(\varepsilon)}
{\left[\varepsilon-\textrm{Re}E_{\sigma}^{\mathrm{edge}}(k_x,\varepsilon)\right]^2+\Gamma(\varepsilon)^2},
\end{equation}
where $\Gamma(\varepsilon)$ is a damping factor for quasiparticle edge states
\begin{equation}
\label{eqRnD:16}
\Gamma(\varepsilon)=\textrm{Im}\Sigma_0(\varepsilon)-\dfrac{D}{B}\textrm{Im}\Sigma_z(\varepsilon).
\end{equation}

Figure~\ref{Fig:3} shows the evolution of total spectral function determined as the sum $A(k,\varepsilon)$ in Eq.~(\ref{eqA1:9}) and $A^{\mathrm{edge}}(k,\varepsilon)$ in Eq.~(\ref{eqRnD:15}) in the prototype 6~nm wide HgTe/Cd$_{0.7}$Hg$_{0.3}$Te QW at several values of the short-range disorder strength $W$. It is seen that although the spectral function broadens with increasing disorder strength, it still clearly represent the quasiparticle bulk dispersion. Interestingly, the bulk quasiparticles at the gapless case at $W=W_c$ mimics massless Dirac fermions as it is in the ``clean'' limit~\cite{QP44b}. If the strength of disorder exceeds a critical value $W_c$, which leads to $\mathbb{Z}_2(\varepsilon)=1$, the bulk states coexist with a pair of quasiparticle helical edge states -- see Fig.~\ref{Fig:3}(c).

Importantly, since $\textrm{Im}\Sigma_z(\varepsilon)$ and $\textrm{Im}\Sigma_0(\varepsilon)$ both vanish in the band-gap region (see Fig.~\ref{Fig:2}(b,c)), the damping factor $\Gamma(\varepsilon)$ turns to zero. As a result, the spectral function of the edge states in the band-gap takes the form of a delta function
\begin{equation}
\label{eqRnD:17}
A^{\mathrm{edge}}(k_x,\varepsilon)=\sum_{\sigma}\delta\left\{\varepsilon-\textrm{Re}E_{\sigma}^{\mathrm{edge}}(k_x,\varepsilon)\right\}.
\end{equation}
The latter means that the edge quasiparticles \emph{do not decay}. The absence of energy dissipation for the quasiparticle helical edge states in the band-gap explains the quantized values of the longitudinal conductance in the presence of an external bias, found previously in numerical calculations in strip geometry~\cite{QP16,QP17}. The latter can be shown explicitly by means of Landauer-B\"{u}ttiker formalism, taking into account that the transmission coefficients along the edge channel do not change~\cite{QP58,QP59}.

Finally, we note the importance of taking into account the dependence of the self-energy matrix on $\varepsilon$ for the description of topological phase transitions. In the case any interactions, and exemplary provided by short-range electrostatic disorder above, our quasiparticle concept goes beyond the previous ``topological Hamiltonian'' $H_{t}(\mathbf{k})$ approach as used by Z.~Wang and collaborators~\cite{QP26,QP27,QP28}.

\section{\label{Sec:Sum} Summary and Notes}
We have presented a general recipe to describe topological phase transitions in condensed matter systems with interactions. By using the simplest BHZ model with the short-range disorder, we have directly demonstrated that that topological invariants in the presence of interactions can be efficiently calculated by means of a non-Hermitian quasiparticle Hamiltonian introduced on the basis of the Green's function. The quasiparticle approach allows us to explicitly demonstrate that the quantized values of the longitudinal conductance found previously in numerical calculations~\cite{QP16,QP17} are due to the vanishing of the damping factor of quasiparticle edge states in a certain range of the Fermi energy and the disorder strength.

Note that despite the relatively simple case of a 2D system with the short-range disorder considered in this paper, the quasiparticle approach should be applicable in general for topological characterization of arbitrary systems (including the ones described beyond the BHZ model~\cite{QP18c}) with an arbitrary type of interaction. Since the latter is the source of the non-Hermitian nature of the quasiparticle Hamiltonian, the presence of interactions can lead to topological phase transitions described by invariants absent in Hermitian systems~\cite{QP25b,QP36c,QP37,QP38,QP54}, including those associated with higher-order topology~\cite{QP42,QP60,QP61}. Moreover, for some types of interactions resulting in specific forms of non-Hermitian quasiparticle Hamiltonians, the conventional bulk-edge correspondence can break down~\cite{QP36o,QP36,QP36b,QP36c}. All these points are missed if one characterizes a condensed matter system in the presence of interactions by means of ``topological Hamiltonian'' $H_{t}(\mathbf{k})$ defined by Eq.~(\ref{eq:2}).

\begin{acknowledgments}
This work was supported by the Occitanie region through the programs ``Terahertz Occitanie Platform'' and ``Quantum Technologies Key Challenge'' (TARFEP project), and by the Elite Network of Bavaria within the graduate program ``Topological Insulators''. We also acknowledge financial support from the French Agence Nationale pour la Recherche through ``Cantor'' (ANR-23-CE24-0022) project and the DFG through the W\"{u}rzburg-Dresden Cluster of Excellence on Complexity and Topology in Quantum Matter -- ct.qmat (EXC 2147, project-id 390858490).
\end{acknowledgments}

\appendix
\section{\label{sec:AppChern} Chern number for two-band non-Hermitian Hamiltonian}
For the case of a two-band system, the calculation of the Chern number by means of Eq.~(\ref{eqA2:9}) can be significantly simplified. Indeed, the integrand in Eq.~(\ref{eqA2:9}) can be calculated relatively easily using the projection operator $\widehat{P}_{s}(\mathbf{k},\varepsilon)=|\Psi_s^{R}(\mathbf{k},\varepsilon)\rangle\langle\Psi_s^{L}(\mathbf{k},\varepsilon)|$:
\begin{multline}
\label{eq:B0}
\nabla_{\mathbf{k}}\times\langle\Psi_s^{L}(\mathbf{k},\varepsilon)|\nabla_{\mathbf{k}}\Psi_s^{R}(\mathbf{k},\varepsilon)\rangle
\cdot{\hat{z}_0}\\
=\epsilon_{ij}\mathrm{Tr}\left[\widehat{P}_{s}(\mathbf{k},\varepsilon)\left(\partial_{i}\widehat{P}_{s}(\mathbf{k},\varepsilon)\right)\left(\partial_{j}\widehat{P}_{s}(\mathbf{k},\varepsilon)\right)\right].
\end{multline}
We remind that here, $s$ is the band index that labels different eigenstates; $\epsilon_{ij}=-\epsilon_{ji}$ denotes the Levi-Civita symbol in two dimensions and the summation over $i$ and $j$ is implied.

In order to find the projection operator explicitly, let us represent an arbitrary two-band non-Hermitian Hamiltonian in the form
\begin{equation}
\label{eq:B1}
\mathbb{H}(\mathbf{k},\varepsilon)=h_{0}(\mathbf{k},\varepsilon)+\mathbf{h}(\mathbf{k},\varepsilon)\cdot\vec{\sigma},
\end{equation}
where $h_{0}$ and $\mathbf{h}=(h_x,h_y,h_z)$ are complex functions of $\mathbf{k}$ and $\varepsilon$ (cf. Eq.~(\ref{eqA2:2})). The eigenvalues of $\mathbb{H}(\mathbf{k},\varepsilon)$ are written as
\begin{equation}
\label{eq:B2}
E_{\pm}(\mathbf{k},\varepsilon)=h_{0}(\mathbf{k},\varepsilon)\pm{\lambda(\mathbf{k},\varepsilon)},
\end{equation}
where $\lambda(\mathbf{k},\varepsilon)$ is defined as
\begin{equation}
\label{eq:B3}
\lambda(\mathbf{k},\varepsilon)=\sqrt{h_x^2(\mathbf{k},\varepsilon)+h_y^2(\mathbf{k},\varepsilon)+h_z^2(\mathbf{k},\varepsilon)}.
\end{equation}
Here, a complex square root should be understood as:
\begin{multline*}
\sqrt{u+iv}=\\
\pm\left(\sqrt{\dfrac{u+\sqrt{u^2+v^2}}{2}}+i\cdot\mathrm{sgn}\left(v\right)\sqrt{\dfrac{-u+\sqrt{u^2+v^2}}{2}}\right).
\end{multline*}
Importantly, by choosing the positive sign in the formula above, it can be shown that by the values of the complex root in Eq.~(\ref{eq:B3}) at $h_x(\mathbf{k},\varepsilon)=0$ and $h_y(\mathbf{k},\varepsilon)=0$, one should mean
\begin{equation}
\label{eq:B4}
\sqrt{h_z^2(\mathbf{k},\varepsilon)}=h_z(\mathbf{k},\varepsilon)\mathrm{sgn}\left[\textrm{Re}\left\{h_z(\mathbf{k},\varepsilon)\right\}\right]
\end{equation}
if $\textrm{Re}\left\{h_z(\mathbf{k},\varepsilon)\right\}{\neq}0$.

Noting that $\mathcal{H}_{qp}^{\uparrow}(\mathbf{k},\varepsilon)$ can be represented as
\begin{multline}
\label{eq:B5}
\mathbb{H}(\mathbf{k},\varepsilon)=E_{+}(\mathbf{k},\varepsilon)\dfrac{1+\mathbf{\hat{h}(\mathbf{k},\varepsilon)}\cdot\vec{\sigma}}{2}~\\
+E_{-}(\mathbf{k},\varepsilon)\dfrac{1-\mathbf{\hat{h}(\mathbf{k},\varepsilon)}\cdot\vec{\sigma}}{2},
\end{multline}
where $\mathbf{\hat{h}}(\mathbf{k},\varepsilon)=\mathbf{h}(\mathbf{k},\varepsilon)/\lambda(\mathbf{k},\varepsilon)$, the projection operator is written as
\begin{equation}
\label{eq:B6}
\widehat{P}_{s}(\mathbf{k},\varepsilon)=\dfrac{1+s\mathbf{\hat{h}(\mathbf{k},\varepsilon)}\cdot\vec{\sigma}}{2}
\end{equation}
with $s={\pm}1$.
Substituting Eq.~(\ref{eq:B6}) into Eq.~(\ref{eq:B0}), the straightforward calculations give
\begin{multline}
\label{eq:B7}
\epsilon_{ij}\mathrm{Tr}\left[\widehat{P}_{s}(\mathbf{k},\varepsilon)\left(\partial_{i}\widehat{P}_{s}(\mathbf{k},\varepsilon)\right)\left(\partial_{j}\widehat{P}_{s}(\mathbf{k},\varepsilon)\right)\right]\\
=\dfrac{i}{2}\dfrac{s}{\lambda(\mathbf{k},\varepsilon)^3}\det\left|
\begin{pmatrix}
h_x(\mathbf{k},\varepsilon) & h_y(\mathbf{k},\varepsilon) & h_z(\mathbf{k},\varepsilon)  \\[4pt]
\dfrac{\partial{h_x(\mathbf{k},\varepsilon)}}{\partial{k_x}} & \dfrac{\partial{h_y(\mathbf{k},\varepsilon)}}{\partial{k_x}} & \dfrac{\partial{h_z(\mathbf{k},\varepsilon)}}{\partial{k_x}}  \\[8pt]
\dfrac{\partial{h_x(\mathbf{k},\varepsilon)}}{\partial{k_y}} & \dfrac{\partial{h_y(\mathbf{k},\varepsilon)}}{\partial{k_y}} & \dfrac{\partial{h_z(\mathbf{k},\varepsilon)}}{\partial{k_y}}
\end{pmatrix}\right|.
\end{multline}
The latter allows one to rewrite the Chern number in Eq.~(\ref{eqA2:9}) in the main text as
\begin{equation}
\label{eq:B8}
C^{(s)}_{\uparrow}(\varepsilon)=-\dfrac{s}{4\pi}{\int}
\dfrac{\mathbf{h}(\mathbf{k},\varepsilon)}{\lambda(\mathbf{k},\varepsilon)^3}\cdot\left[\nabla_{k_x}\mathbf{h}(\mathbf{k},\varepsilon)\times
\nabla_{k_y}\mathbf{h}(\mathbf{k},\varepsilon)\right]
{d^2\textbf{k}}.
\end{equation}

\section{\label{sec:AppRot} Axial rotation symmetry and SCBA}
Due to the axial rotational symmetry of $\mathcal{H}_{0}(\mathbf{k})$ in Eq.~(\ref{eqA2:1}), its wave-function can be presented in the form:
\begin{equation}
\label{eq:A1}
\Psi_{0}(\mathbf{k})=U(\theta_{\mathbf{k}})^{-1}\Psi_{0}(k),
\end{equation}
where $k=|\textbf{k}|$, $k_x=k\cos\theta_{\mathbf{k}}$, $k_y=k\sin\theta_{\mathbf{k}}$, and
\begin{equation}
\label{eq:A2}
U(\theta)=\begin{pmatrix}
e^{i\theta/2} & 0 & 0 & 0 \\
0 & e^{i3\theta/2} & 0 & 0 \\
0 & 0 & e^{-i\theta/2} & 0 \\
0 & 0 & 0 & e^{-i3\theta/2}\end{pmatrix}.
\end{equation}
Therefore, the Green's function in Eq.~(\ref{eqA1:2}) can be presented in the form
\begin{equation}
\label{eq:A3}
\hat{G}(\mathbf{k},\varepsilon)=U(\theta_{\mathbf{k}})\hat{G}(k,\varepsilon)U(\theta_{\mathbf{k}})^{-1},
\end{equation}
with
\begin{equation}
\label{eq:A4}
\hat{G}(k,\varepsilon)=\left[\varepsilon-\tilde{\mathcal{H}}_{0}(k)-\hat{\Sigma}(k,\varepsilon)\right]^{-1},
\end{equation}
which depends only on $k$. This shows that $\hat{G}(\mathbf{k},\varepsilon)$ depends on the angle via the terms of $U(\theta_{\mathbf{k}})$. We note that $\tilde{\mathcal{H}}_{0}(k)$ differs from $\mathcal{H}_{0}(\mathbf{k})$ by
\begin{widetext}
\begin{equation}
\label{eq:A5}
\tilde{\mathcal{H}}_{0}(k)=U(\theta_{\mathbf{k}})\mathcal{H}_{0}(\mathbf{k})U(\theta_{\mathbf{k}})^{-1}
=\begin{pmatrix}
d_{0}(k)+d_z(k)\sigma_z+Ak\sigma_x & 0 \\ 0 & d_{0}(k)+d_z(k)\sigma_z-Ak\sigma_x\end{pmatrix}.
\end{equation}

By using Eq.~(\ref{eq:A3}), we have
\begin{equation}
\label{eq:A6}
\hat{\Sigma}(\textbf{k},\varepsilon)=n_{i}U(\theta_{\mathbf{k}})\int\dfrac{d^2\textbf{k}^\prime}{(2\pi)^2}
\tilde{v}(\textbf{k}-\textbf{k}^\prime)U(\theta_{\mathbf{k}^\prime}-\theta_{\mathbf{k}})\hat{G}(k^\prime,\varepsilon)U(\theta_{\mathbf{k}^\prime}-\theta_{\mathbf{k}})^{-1}\tilde{v}(\textbf{k}^\prime-\textbf{k})U(\theta_{\mathbf{k}})^{-1}.
\end{equation}
Thus, similar to Eq.~(\ref{eq:A3}), the self-energy matrix can be written as
\begin{equation}
\label{eq:A7}
\hat{\Sigma}(\mathbf{k},\varepsilon)=U(\theta_{\mathbf{k}})\hat{\Sigma}(k,\varepsilon)U(\theta_{\mathbf{k}})^{-1},
\end{equation}
where matrix $\hat{\Sigma}(k,\varepsilon)$ has a form
\begin{equation}
\label{eq:A8}
\hat{\Sigma}(k,\varepsilon)=n_{i}\int\limits_0^{K_c}\dfrac{k^\prime dk^\prime}{2\pi}\begin{pmatrix}
V_0(k,k^\prime)^2G_{11}^\prime & V_{+1}(k,k^\prime)^2G_{12}^\prime & 0 & 0\\ V_{-1}(k,k^\prime)^2G_{21}^\prime & V_0(k,k^\prime)^2G_{22}^\prime & 0 & 0\\
0 & 0 & V_0(k,k^\prime)^2G_{33}^\prime & V_{-1}(k,k^\prime)^2G_{34}^\prime \\
0 & 0 & V_{+1}(k,k^\prime)^2G_{43}^\prime & V_0(k,k^\prime)^2G_{44}^\prime \end{pmatrix}.
\end{equation}
\end{widetext}
Here, $G_{ij}^\prime\equiv G_{ij}(k^\prime,\varepsilon)$ are the component of the Green's function in Eq.~(\ref{eq:A4}), and $V_n(k,k^\prime)^2$ is written as
\begin{equation}
\label{eq:A9}
V_n(k,k^\prime)^2=\int\limits_0^{2\pi}\dfrac{d\theta}{2\pi}|\tilde{v}(\textbf{k}-\textbf{k}^\prime)|^2
\cos n\theta.
\end{equation}
In Eq.~(\ref{eq:A8}), we introduce a cut-off wave-vector $K_c=\pi/a_{\mathrm{0}}$ (where $a_{\mathrm{0}}$ is the lattice constant, see Tab.~\ref{tab:1}, cf. Ref.~\cite{QP18b}), which corresponds to the size of the first Brillouin zone.

\begin{table*}
\caption{\label{tab:C2} Several other examples of calculating Chern numbers for certain forms of the self-energy matrix that break rotational symmetry. Here, $\alpha(\varepsilon)$, $\beta(\varepsilon)$, $\Omega(\varepsilon)$ and $\Sigma_{z}^{(0)}(\varepsilon)$ are complex functions of $\varepsilon$.}
\begin{ruledtabular}
\begin{tabular}{c|c|c|c}
$\Sigma_{x}(\mathbf{k},\varepsilon)$ & $\Sigma_{y}(\mathbf{k},\varepsilon)$ & $\Sigma_{z}(\mathbf{k},\varepsilon)$ & $C^{(s)}(\varepsilon)$ \\
\hline
$\Omega(\varepsilon)k_x$ & $\Omega(\varepsilon)k_y$ & $\Sigma_{z}^{(0)}(\varepsilon)+\alpha(\varepsilon){k}_x^2+\beta(\varepsilon){k}_y^2$ &  $-\dfrac{s}{2}\left\{
\int\limits_{0}^{2\pi}\dfrac{\mathrm{sgn}\left[B-\textrm{Re}\left\{
\alpha(\varepsilon)\cos^2(\theta)+\beta(\varepsilon)\sin^2(\theta)\right\}
\right]}{2\pi}
d\theta
+\mathrm{sgn}\left[M+\textrm{Re}\Sigma_z^{(0)}(\varepsilon)\right]\right\}$ \\[11pt]
$\Omega(\varepsilon)k_x$ & 0 &
$\Sigma_{z}^{(0)}(\varepsilon)$ &  $-\dfrac{s}{2}\left\{
\mathrm{sgn}\left[B\right]
+\mathrm{sgn}\left[M+\textrm{Re}\Sigma_z^{(0)}(\varepsilon)\right]\right\}\mathrm{sgn}\left[A\left(A+\textrm{Re}\Omega(\varepsilon)\right)\right]$ \\[11pt]
0 & $\Omega(\varepsilon)k_y$ &
$\Sigma_{z}^{(0)}(\varepsilon)$ &  $-\dfrac{s}{2}\left\{
\mathrm{sgn}\left[B\right]
+\mathrm{sgn}\left[M+\textrm{Re}\Sigma_z^{(0)}(\varepsilon)\right]\right\}\mathrm{sgn}\left[A\left(A+\textrm{Re}\Omega(\varepsilon)\right)\right]$ \\[11pt]
0 & $\Omega(\varepsilon)k_x$ &
$\Sigma_{z}^{(0)}(\varepsilon)$ &  $-\dfrac{s}{2}\left\{
\mathrm{sgn}\left[B\right]
+\mathrm{sgn}\left[M+\textrm{Re}\Sigma_z^{(0)}(\varepsilon)\right]\right\}
\dfrac{\mathrm{sgn}\left[A^2-\left|A\cdot\textrm{Im}\Omega(\varepsilon)\right|\right]+1}{2}$ \\[11pt]
\end{tabular}
\end{ruledtabular}
\end{table*}

\section{\label{sec:AppSigma} Chern number in two-band model with certain self-energy matrices}
In the main text, as an example analytically illustrating the application of the quasiparticle concept, we consider the short-range electrostatic disorder, resulting to the diagonal form of the \emph{isotropic} self-energy matrix independent of energy. Let us now focus on a few examples of \emph{anisotropic} $\hat{\Sigma}(\mathbf{k},\varepsilon)$ as a function of $\mathbf{k}$ (without concretizing the source of the resulted interaction), also allowing analytical calculation of the Chern numbers and the associated $\mathbb{Z}_2$ invariant within the BHZ model. For brevity, we consider the calculations of the Chern number for the upper block of the BHZ Hamiltonian~(\ref{eqA2:1}), omitting the upper arrow index. In the most general case, the upper block of quasiparticle Hamiltonian~(\ref{eqA2:4}) can be represented in the form
\begin{multline}
\label{eqC1:1}
\mathcal{H}_{qp}(\mathbf{k},\varepsilon)=H_{\mathrm{BHZ}}(\mathbf{k})+\Sigma_{0}(\mathbf{k},\varepsilon)~\\
+\Sigma_{x}(\mathbf{k},\varepsilon)\sigma_x+\Sigma_{y}(\mathbf{k},\varepsilon)\sigma_y+\Sigma_{z}(\mathbf{k},\varepsilon)\sigma_z.
\end{multline}
To calculate the Chern number by means of Eq.~(\ref{eq:B8}), one should know an exact form of $\Sigma_{x}(\mathbf{k},\varepsilon)$, $\Sigma_{y}(\mathbf{k},\varepsilon)$ and $\Sigma_{z}(\mathbf{k},\varepsilon)$ as a function of $\mathbf{k}$. Let us assume
\begin{equation}
\label{eqC1:2}
\Sigma_{x}(\mathbf{k},\varepsilon)=\Omega(\varepsilon)k_y,~~~\Sigma_{y}(\mathbf{k},\varepsilon)=0,~~~\Sigma_{z}(\mathbf{k},\varepsilon)=\Sigma_{z}^{(0)}(\varepsilon),
\end{equation}
where $\Omega(\varepsilon)$ and $\Sigma_{z}^{(0)}(\varepsilon)$ are complex functions of $\varepsilon$.

In this case, the Chern number in the polar system is written as
\begin{equation}
\label{eqC1:3}
C^{(s)}(\varepsilon)=-\dfrac{sA^2}{4\pi}\int\limits_{0}^{2\pi}d{\theta}\int\limits_{0}^{\infty}
\dfrac{Bk^2+M+\Sigma_{z}^{(0)}(\varepsilon)}{\Pi({k},\theta,\varepsilon)^{3/2}}k{d{k}},
\end{equation}
where
\begin{multline}
\label{eqC1:4}
\Pi({k},\theta,\varepsilon)=\left[M+\Sigma_z^{(0)}(\varepsilon)-Bk^2\right]^2\\
+k^2\left[A^2+\Omega(\varepsilon)^2\sin^2\theta+A\Omega(\varepsilon)\sin2\theta\right].
\end{multline}
Integration over the wave vector leads to
\begin{multline}
\label{eqC1:5}
C^{(s)}(\varepsilon)=-\dfrac{sA^{2}}{4\pi}\left\{
\mathrm{sgn}\left[B\right]
+\mathrm{sgn}\left[M+\textrm{Re}\Sigma_z^{(0)}(\varepsilon)\right]\right\}\\
\times\int\limits_{0}^{2\pi}\dfrac{d{\theta}}{A^2+\Omega(\varepsilon)^2\sin^2\theta+A\Omega(\varepsilon)\sin2\theta}.
\end{multline}

Replacing the integration over an angle $\theta$ with the integration over the contour of a unit circle in the complex plane results in
\begin{multline}
\label{eqC1:6}
\int\limits_{0}^{2\pi}\dfrac{d{\theta}}{A^2+\Omega(\varepsilon)^2\sin^2\theta+A\Omega(\varepsilon)\sin2\theta}\\
=\dfrac{4i}{\Omega(\varepsilon)\left[\Omega(\varepsilon)+2iA\right]}\oint\limits_{|z|=1}\dfrac{{z}d{z}}{\left(z^2-z_1^2\right)\left(z^2-z_2^2\right)},
\end{multline}
where
\begin{equation}
\label{eqC1:7}
z_1^2=1-\dfrac{2iA}{\Omega(\varepsilon)},~~~~~~~~z_2^2=1-\dfrac{2iA}{\Omega(\varepsilon)+2iA}.
\end{equation}

Representing complex $\Omega(\varepsilon)=\Omega_0(\varepsilon)+i\omega(\varepsilon)$ (where $\Omega_0(\varepsilon)$ and $\omega(\varepsilon)$ are both real) and using the standard residue theory of complex analysis, one can show that the integral over $\theta$ in Eq.~(\ref{eqC1:6}) vanishes if $A^2<|A\omega(\varepsilon)|$ and equals to $2\pi/A^2$ if $A^2>|A\omega(\varepsilon)|$. To sum up all of the above, the Chern number in Eq.~(\ref{eqC1:5}) can be rewritten as
\begin{multline}
\label{eqC1:8}
C^{(s)}(\varepsilon)=-\dfrac{s}{2}\left\{
\mathrm{sgn}\left[B\right]
+\mathrm{sgn}\left[M+\textrm{Re}\Sigma_z^{(0)}(\varepsilon)\right]\right\}\\
\times\dfrac{\mathrm{sgn}\left[A^2-\left|A\cdot\textrm{Im}\Omega(\varepsilon)\right|\right]+1}{2}.
\end{multline}
Thus, the presence of an additional component in $\hat{\Sigma}(\mathbf{k},\varepsilon)$, compared to the case considered in the main text, can indeed lead to a dependence of the Chern numbers and the $\mathbb{Z}_2$ invariant on the imaginary part of the self-energy matrix. Additionally, Table~\ref{tab:C2} summarizes several other momentum-dependent cases for $\Sigma_{x}(\mathbf{k},\varepsilon)$, $\Sigma_{y}(\mathbf{k},\varepsilon)$ and $\Sigma_{z}(\mathbf{k},\varepsilon)$ that also allow the Chern number to be calculated analytically within the BHZ model.

%

\end{document}